%% file: main.tex
\input macros

\input firstpage

\input acknowledgement

\input contents

\input abstract

\input chapters

\input references

\bye

%% file: macros.tex
\input eplain   
\input epsf

\input braket.sty


\magnification\magstep1

\voffset=+0.5cm
\hoffset=+1.1cm
\advance \vsize by -0.5cm
\advance \hsize by -1.1cm

\def\vspace#1{\vskip#1}
\font\twelverm=csr12
\def\large{\twelverm}
\font\twelvebf=csbx12
=\fontname\twelvebf\space scaled\magstep1

\newcount\eqcount
\def\incno#1{\global\advance\eqcount1\definexref{#1}{\the\eqcount}{eq}}
\def\no#1{\incno{#1}\eqno{(\the\eqcount)}}
\def\nno#1{\incno{#1}&{(\the\eqcount)}}
\def\rno#1{(\refn{#1})}

\newcount\secno
\newcount\subno
\def\secskip{\vskip 18pt plus 6pt minus 6pt}

=\fontname\tenbf\space scaled\magstep2
=\fontname\tenbf\space scaled\magstep1
=\fontname\tenbf\space scaled\magstep0

\def\title#1{\centerline{\titlefont #1}\bigskip}
\def\sec#1\par{\advance\secno by1\subno=0\secskip\goodbreak\noindent%
{\subfont \the\secno. #1}%
\writenumberedtocentry{sec}{#1}{\the\secno.}\medskip\nobreak\noindent}
\def\sub#1\par{\advance\subno by 1\medskip\goodbreak\noindent%
{\edef\n{\the\secno.\the\subno}{\subsubfont\n\ #1}%
\writenumberedtocentry{sub}{#1}{\n}}\par\nobreak\noindent}

\newcount\figno
\def\dref#1{\global\advance\figno by 1\definexref{#1}{\the\figno}{}\the\figno}

\font\twelverm=csr12
\def\d{{\rm d}}

\def\psfigw#1#2#3#4{\medskip\noindent\hfill\epsfxsize=#4\epsfbox{#1}\hfill\break
\centerline{Fig. \dref{#2}: #3}\medskip}

\def\chapter#1\par{\advance\secno by1\subno=0\secskip\goodbreak\vfill\eject\noindent%
{\Large Chapter \the\secno\bigskip\noindent#1}%
\writenumberedtocentry{sec}{#1}{\the\secno}\medskip\nobreak\noindent}

\def\section#1\par{\advance\subno by 1\bigskip\goodbreak\noindent%
{\edef\n{\the\secno.\the\subno}{\twelvebf\n\ #1}%
\writenumberedtocentry{sub}{#1}{\n}}\smallskip\nobreak\noindent}

\def\unnumberedchapter#1{\secskip\goodbreak\vfill\eject\noindent%
{\Large #1}%
\writetocentry{unnumbered}{#1}\medskip\nobreak\noindent}

\def\unnumberedchapter#1{\secskip\goodbreak\vfill\eject\noindent%
{\Large #1}%
\writetocentry{unnumbered}{#1}\medskip\nobreak\noindent}

\def\references{\unnumberedchapter{References}}
\def\contents{\noindent{\Large Contents}\bigskip\nobreak\noindent}

\def\pagenumbers{\footline={\hss\tenrm\folio\hss}}

\def\url#1{{\tt #1}}

\def\back{}
\def\col#1#2{\left(\matrix{#1\cr#2\cr}\right)}
\def\row#1#2{\left(\matrix{#1&#2\cr}\right)}
\def\mat#1#2#3#4{\left(\matrix{#1&#2\cr#3&#4\cr}\right)}
\def\matd#1#2{\left(\matrix{#1&\back0\cr0&\back#2\cr}\right)}
\def\p#1#2{{\partial#1\over\partial#2}}
\def\cg#1#2#3#4#5#6{({#1},\,{#2},\,{#3},\,{#4}\,|\,{#5},\,{#6})}
\def\half{{\textstyle{1\over2}}}
\def\jsym#1#2#3#4#5#6{\left\{\matrix{
{#1}&{#2}&{#3}\cr
{#4}&{#5}&{#6}\cr
}\right\}}

\font\dsrom=dsrom10

\font\mib=cmmib10
\def\balpha{\hbox{\mib\char"0B}}

\def\bsigma{\hbox{\mib\char"1B}}

%

\font\mibsmall=cmmib7
\def\bsigmasmall{\hbox{\mibsmall\char"1B}}

%% file: firstpage.tex
\nopagenumbers
\vspace{15mm}

\centerline{\large Charles University in Prague}
\centerline{\large Faculty of Mathematics and Physics}

\vspace{5mm}

\centerline{\Large BACHELOR THESIS}

\vspace{10mm}

\centerline{\epsfxsize=0.28\hsize\epsfbox{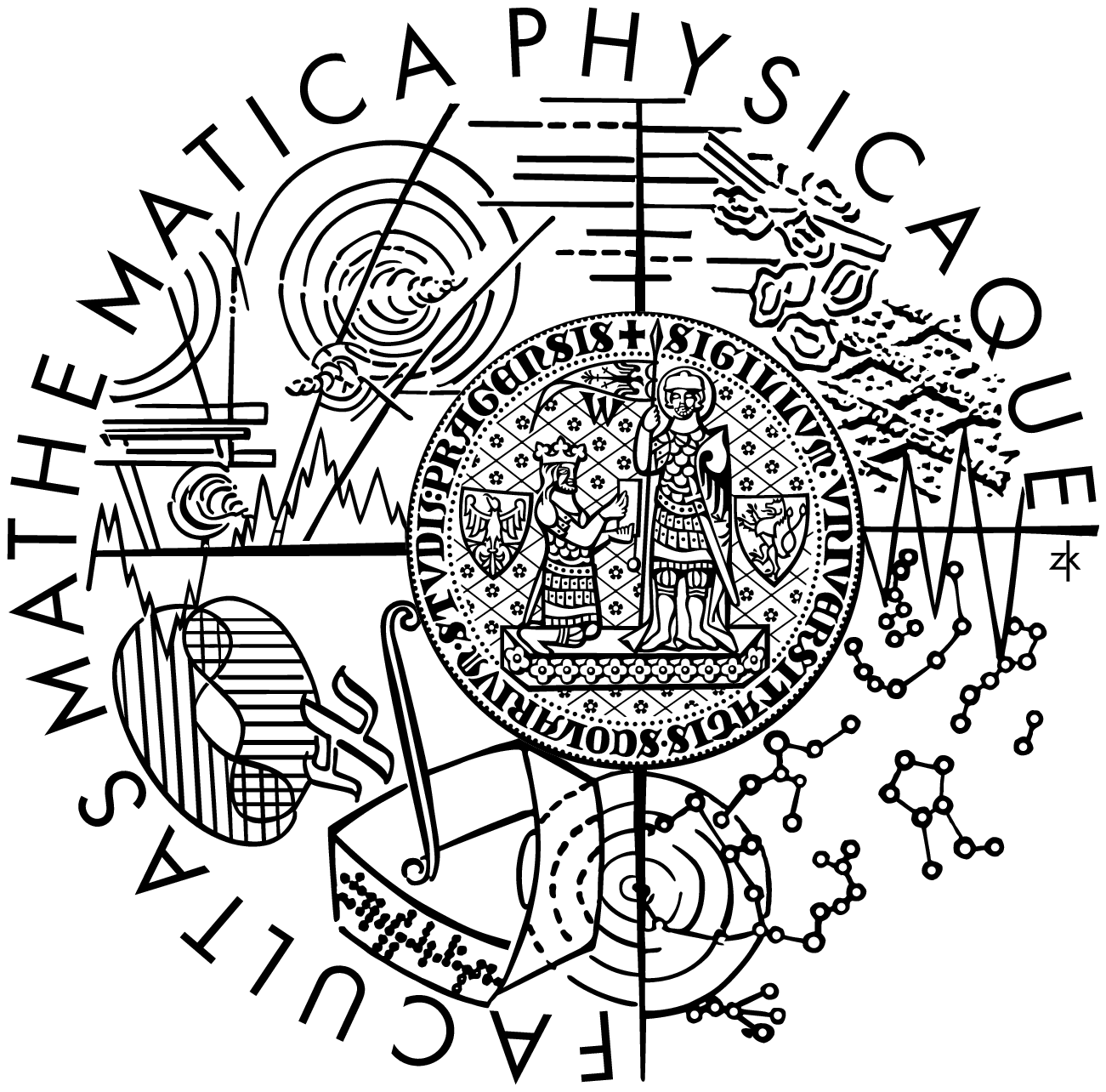}}

\vspace{15mm}

\centerline{\large Ondøej Èertík}
\vspace{5mm}
\centerline{\Large Numerical solution of the radial Dirac equation}
\centerline{\Large in pseudopotential construction}
\vspace{5mm}
\centerline{\large Institute of Theoretical Physics}
\vspace{15mm}

\hfil\vbox{
\hbox{\large Supervisor: RNDr. Jiøí Vackáø, CSc., Institute of Physics,}
\hbox{\large \hskip2cm Academy of Sciences, Czech Republic}
\hbox{\large Field of study: general physics}
}\hfil

\vspace{40mm}

\centerline{\large 2006}

\eject

%% file: acknowledgement.tex
\vspace{10mm} 


\noindent{\large I would like to thank RNDr. Jiøí Vackáø, CSc., for many
discussions, enlightening a lot of issues in quantum mechanics to me
and for his time helping me with this work. I also thank RNDr. Ondøej ©ipr,
CSc., for explaining me several theoretical problems.}

\vfill 


\noindent{\large I declare that I wrote the thesis by myself and listed all
used sources. I agree with making the thesis publicly available.}

\bigskip
\noindent{\large Prague, June 3, 2006\hfill Ondøej Èertík}

\eject

%% file: contents.tex
\pagenumbers

\contents

\readtocfile

\vfill
\eject

%% file: abstract.tex
\noindent
{\bf Title}: Numerical solution of the radial Dirac equation 
in pseudopotential construction\hfill\break
{\bf Author}: Ondøej Èertík\hfill\break
{\bf Department}: Institute of Theoretical Physics\hfill\break
{\bf Supervisor}: RNDr. Jiøí Vackáø, CSc., Institute of Physics, ASCR\hfill\break
{\bf Supervisor's e-mail}: {\tt vackar@fzu.cz}\hfill\break

\noindent {\bf Abstract}: In the present work we study numerical solution of
the radial Dirac equation in a specific case --- {\it ab-initio}
pseudopotential generating process --- which is needed within the electronic
structure calculations using a Density Functional Theory (DFT)
combined with a pseudopotential method. We give a brief
introduction to DFT, derive the radial Dirac and Schr\"odinger
equations, show how to solve them both for a given energy and as an eigenvalue
problem using a known asymptotic behavior of the solution. Next we compare the
nonrelativistic and relativistic eigenvalues for one electron atom. Finally
we state a few words about the computer implementation.

\bigskip

\vspace{10mm}






\vfill
\eject

%% file: chapters.tex








\input intro

\input chapter1

\input chapter2

\input chapter3

\input chapter4

\input chapter5

\input conclusion

\input appendix

%% file: intro.tex
\def\em{\it}

\chapter{Introduction}

The outcome of this thesis is an algorithm solving the radial Dirac equation
(together with the corresponding computer code) in a specific case, as a part 
of a particular method for electronic structure calculations in solid-state
physics.

The electronic structure calculations are essential for any theoretical study
of materials, which is currently an extremely active field of research 
and is often denoted as {\em computational material science}.
In past few decades, electronic structure calculations made a significant 
contribution to our understanding of material properties,
using a large variety of continuously developing methods and approaches.

This work is related to one of up-to-date {\em ab-initio pseudopotential} 
methods based on the {\em density-functional theory} 
(DFT)\cite{wikidft,pickett},
particularly to the pseudopotential generating process within the 
{\em all-electron pseudopotential} method
\cite{vackarAEPP1,vackarAEPP2}.

{\em Ab-initio method} means that the method do not require any empirical 
parameters as an input to the calculation --- being based on equations 
derived directly from theoretical principles, with no need for 
experimental input data.

The standard result of the particular calculation within DFT is the total 
energy of the given system and the electronic charge density (or the wave 
function of electronic states that, integrated over the space and occupied 
states, form the charge density). Within DFT, the electronic charge density 
is a key quantity containing complete information about the system. 
This information allows us to answer almost any question we might ask about 
the solid: 
the structure of electronic states provides information about thermal and
electrical conductivity, e.g. metallic or semiconductive behavior, 
optical or X-ray emission/absorption spectra and scattering properties;
minimization of the total energy with respect to atomic positions provides 
an equilibrium geometry; derivatives of the total energy gives bulk modulus 
and elastic constants; spin structure of electronic states and derivatives 
of the total energy with respect to an external fields can provide dielectric
or magnetic properties of the solid, etc.

At present, several more-or-less ab-initio computational methods withing DFT 
are widely used in solid state physics. Every method usually has it's own domain 
in which it excels. The success of a method is determined by many factors, 
including its usage in particular computer codes and their efficiency, 
accuracy etc. And new methods can emerge in the future. It is not the
aim of this thesis to compare all the methods available. It will suffice 
to say that in present, the majority of calculations are based on the
local-density approximation (or its extensions) to DFT, which 
we review very briefly now\cite{pickett}.
It leads to the following Kohn-Sham equations (for $N$ electrons):
$$\left\{ -{1\over2m}\nabla^2+
\sum_{n=1}^N V_n^{\rm ion}({\bf x}-{\bf X}_n)+
V_H({\bf x})+V_{XC}({\bf x})
\right\}\psi_i({\bf x})=\epsilon_i\psi_i({\bf x})\,,\no{kohn-sham}$$
where the charge density is
$$\rho=\sum_{i=1}^N |\psi_i|^2\,,$$
the Hartree potential $V_H$ is given by
$$\nabla^2V_H=-4\pi\rho\,,$$
and the exchange-correlation potential $V_{XC}$ is just a given function of the
charge density $\rho$. The $V_n^{\rm ion}$ is the electrostatic 
potential of atomic nuclei. Within the family of up-to-date
pseudopotential methods, it is constructed from 
nonlocal pseudopotentials. There are several ways how to do that. 

These equations need to be solved self-consistently, so that the charge
density, which is used for the construction of the Hartree and
exchange-correlation potential is the same as the charge density calculated from
the equations, see also fig. \refn{schema01}.

\psfigw{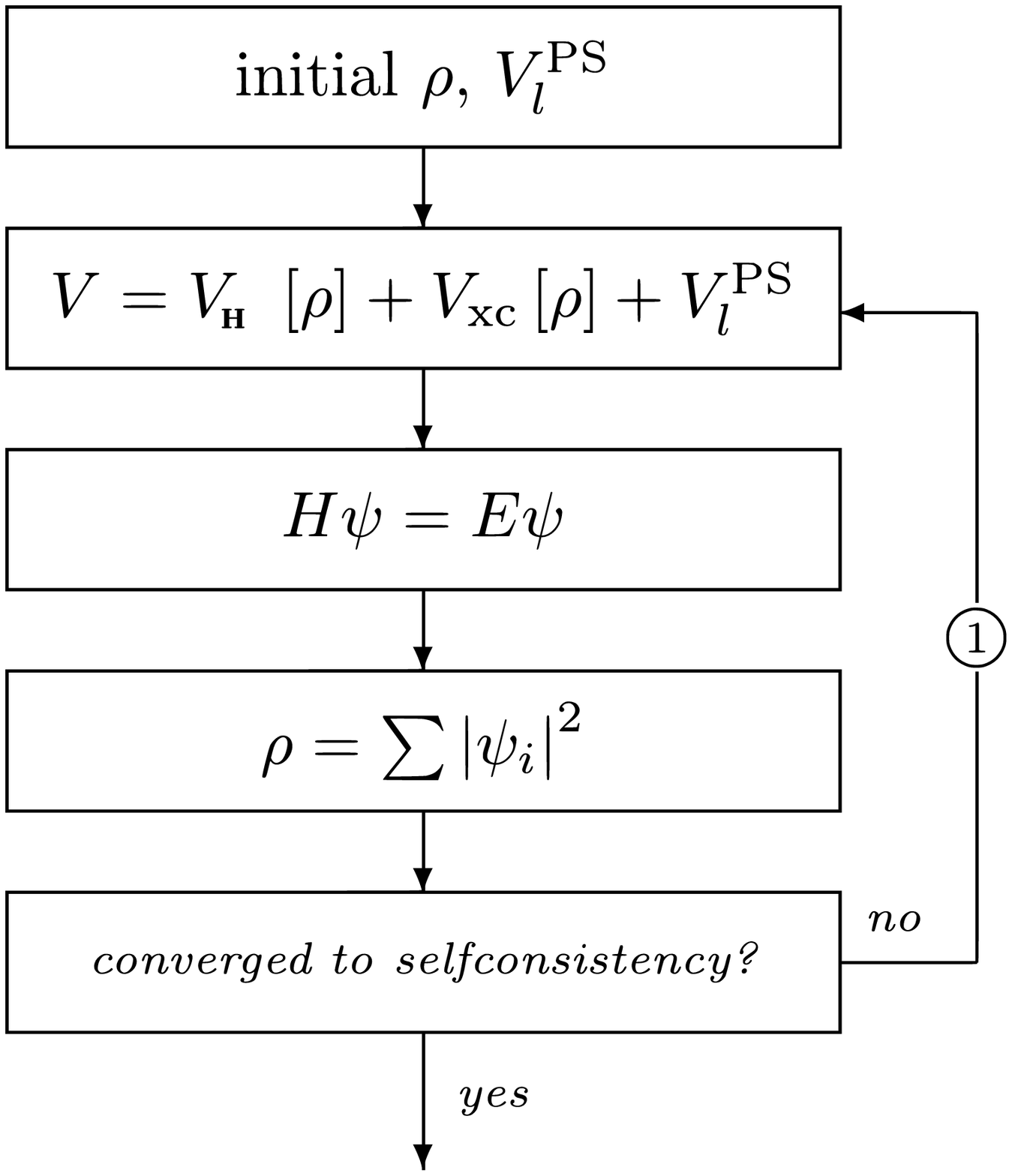}{schema01}{Self-consistency cycle}{6cm}

Numerous methods have been developed to solve the
resulting system of one-particle Kohn-Sham equations \rno{kohn-sham}. 
Depending on the basis set used and other features of the particular method, 
some of them (in fact: almost all of them) need the radial part of the equation
to be solved. 
For heavy atoms (approximately, starting at atomic numbers around 40-50) 
the equations \rno{kohn-sham} are considerably inaccurate due to
the non-negligible relativistic effects, especially for the core electronic states
bounded with relatively high energies. 
For this reason, it is desirable to use a relativistic Dirac
equation instead, which is the aim of this thesis.

In the present thesis we first derive and show how to solve the radial Schr\"odinger
equation for a given energy. Then we move to the relativistic case, Dirac
equation, and do the same. Next we show how to solve the eigenproblem, that is,
how to determine the energies, for which the solution is normalizable. Then we
say something about the computer implementation together with some results of
the program. In the appendix we define atomic units (which is sometimes a
little confusing issue) and we derive spin-angular functions including some of their
properties we need in our work.

%% file: chapter1.tex
\chapter{Schr\"odinger Equation}

\section{Introduction}

The Schr\"odinger equation describes a nonrelativistic particle in a potential
field. It cannot be derived, we always have to postulate something, more or
less equivalent to the equation itself:
$$H\psi=E\psi\,,$$
$$H={p^2\over2M}+V\,,\no{schrodinger}$$
where $\psi(x,y,z)$ is a (complex) function, ${\bf p}=-i\hbar\nabla$,
$M$ is a mass of the particle (in our case an electron), $V(x,y,z)$ the
potential field (in our case we only have a spherically symmetric
field $V(r)$). A physical quantity we are actually interested in
is a probability density $\rho=\psi^*\psi$. 

The electron has to be somewhere in the universe, thus we want 
$$\int \psi\,\d^3x=1\,.\no{normalization}$$
For a given energy $E$, we can always solve the equation, but for the energies
not lying in the spectrum of $H$, the solution exponentially diverges to
infinity and such a solution cannot be normalized so that \rno{normalization}
holds. To be precise - there actually exist physical solutions, which are not
normalizable according to \rno{normalization} (the ones lying in the continuous
part of the spectrum, for example a free electron, $V=0$), but in our case of
bounded states in a potential, we always have a discrete spectrum.

The condition \rno{normalization} picks up only certain energies
(eigenvalues), when the solution doesn't diverge. We label the
energies by an integer number $n$ starting from the lowest one $n=1$, second
lowest $n=2$ etc. 

Besides energy, the solution also depends on quadrate of angular momentum ($l$)
and it's $z$ component ($m$). As it turns out, the radial part of the
solution depends on $n$ and $l$ only.

So we want to solve the eigenvalue problem of finding a solution for a given
$n$ and $l$.

\section{Radial Schr\"odinger equation}

We have a spherically symmetric potential energy
$$V({\bf x})=V(r)\,.$$
State with a given square of an angular momentum (eigenvalue
$l(l+1)$) and its $z$ component (eigenvalue $m$) is described by the wave
function
$$\psi_{nlm}({\bf x})=R_{nl}(r)\,Y_{lm}\left({\bf x}\over r\right)\,,\no{psi}$$
where $R_{nl}(r)$ obeys the equation \cite{formanek} (eq. 2.400)
$$R_{nl}''+{2\over r}R_{nl}'+{2M\over\hbar^2}(E-V)R_{nl}-
{l(l+1)\over r^2}R_{nl}=0\,.\no{radial}$$
This is called the radial Schr\"odinger equation which
we want to solve numerically.

The derivation is well-known \cite{formanek,sakurai}, so just briefly.
Basically, it's just a separation of variables: we decompose the space as a
tenzor product $\hbox{\dsrom R}^3=\hbox{\dsrom S}^2\times\hbox{\dsrom R}$,
where $\hbox{\dsrom S}^2$ is a unit sphere and $\hbox{\dsrom R}$ is the radial
part. We choose a basis in $\hbox{\dsrom S}^2$, it turns out that spherical
harmonics $Y_{lm}$ are a good choice as they are eigenvectors of $L^2$ and
$L_3$. We will search for all solutions of the form \rno{psi}. Substituting
\rno{psi} into the equation \rno{schrodinger} yields \rno{radial}: the trick is
to write $\nabla^2$ in spherical coordinates, the angular derivatives will
then act on $Y_{lm}$ only, thus separating the equaion. It turns out, that all
the solutions $R_{nl}$ form a basis of $\hbox{\dsrom R}$. So we have found a
basis of $\hbox{\dsrom R}^3$, which is also a solution of \rno{schrodinger} and
thus any other solution can be found as a (possibly infinite) linear
combination of $\psi_{nlm}$.

\section{Numerical integration for a given $E$}

Equation \rno{radial} is the linear ordinary differential equation of the second
order, so the general solution is a linear combination of two independent
solutions. Normally, the $2$ constants are determined from initial and/or
boundary conditions. In our case, however, we don't have any other condition
besides being interested in solutions that we can integrate on the interval
$(0,\infty)$ (and which are normalizable), more exactly we want
$R\in L^2$ and $\int_0^\infty r^2R^2\,\d r=1$. 

It can be easily shown by a direct substitution, that there are only two
asymptotic behaviors near the origin: $r^l$ and $r^{-l-1}$. We are interested
in quadratic integrable solutions only, so we are left with $r^l$
and only one integration constant, which we calculate from a normalization.
This determines the solution uniquely.

All the integration algorithms needs to evaluate $R''$, which is a
problem at the origin, where all the terms in the equation are infinite,
although their sum is finite. We thus start to integrate the equation at some
small $r_0$ (for example $r_0=10^{-10}\rm\,a.u.$), where all the terms in the
equation are finite. If we find the initial conditions $R(r_0)$ and
$R'(r_0)$, the solution is then fully determined.

If $r_0$ is sufficiently small, we can set $R(r_0)=r_0^l$ and
$R'(r_0)=lr_0^{l-1}$. This works fine for $l>0$. For $l=0$, it is not strictly
correct, but it works well
in practice because the fourth-order Runge-Kutta method is able to quickly
correct the initial derivative guess.

So when somebody gives us $l$ and $E$, we are now able to compute the solution
but the multiplicative constant that is later determined from a
normalization. As was already mentioned, we used the fourth-order Runge-Kutta
method that proved very suitable for this problem.

\section{Asymptotic behavior}

The asymptotic behavior is important for the integration routine to find
the correct solution for a given $E$. In this section we look into more details
of the asymptotic expansion and illustrate it on 2 examples.

It is well known, that the first
term of the Taylor series of the solution is $r^l$, independent
of the potential \cite{formanek} (eq. 2.408). This is enough
information to find the correct solution for $l>0$ because the only
thing we need to know is the value of the wave function and its derivative
near the origin, which is effectively $r_0^l$ and $lr_0^{l-1}$ for some small
$r_0$. The problem is with $l=0$, where the
derivative cannot be calculated just from $l$ and $r_0$. This section shows why
and in the next section we show how we solved the problem.

We start with the radial Schr\"odinger equation \rno{radial}
and we shall search for the solution $R$ in the form of a Taylor series:
$$R=a_0+a_1r+a_2r^2+\dots=\sum_{k=0}^\infty a_kr^k\,.$$
Substituting this into the equation we get:
$$
\sum_{k=0}^\infty r^ka_k\left[k(k+1)-l(l+1)\right]+
{2M\over\hbar^2}(E-V)\sum_{k=2}^\infty r^ka_{k-2}
=0\,.\no{asym1}
$$
Let's assume we have a potential $V$ of the form:
$$V={v_{-1}\over r}+v_0+v_1r+v_2r^2+\dots=\sum_{j=-1}^\infty v_jr^j\,,$$
we rearrange the double sum on the right hand side of \rno{asym1}
$$
V\sum_{k=2}^\infty r^ka_{k-2}=
\sum_{j=-1}^\infty\sum_{k=2}^\infty v_jr^jr^ka_{k-2}=
\sum_{j=-1}^\infty\sum_{k=j+2}^\infty v_jr^ka_{k-j-2}=
$$
$$
=\sum_{j=0}^\infty\sum_{k=j}^\infty v_{j-1}r^{k+1}a_{k-j}=
\sum_{k=0}^\infty\sum_{j=0}^k v_{j-1}r^{k+1}a_{k-j}=
\sum_{k=0}^\infty r^{k+1}\sum_{j=0}^k v_{j-1}a_{k-j}=
$$
$$
=\sum_{k=1}^\infty r^{k}\sum_{j=0}^{k-1} v_{j-1}a_{k-j-1}
$$
So we get:
$$\sum_{k=0}^\infty r^ka_k\left[k(k+1)-l(l+1)\right]+
{2M\over\hbar^2}E\sum_{k=2}^\infty r^ka_{k-2}+
$$
$$
-{2M\over\hbar^2}\sum_{k=1}^\infty r^{k}\sum_{j=0}^{k-1} v_{j-1}a_{k-j-1}=0\,.
$$
This equation holds for every $r$, thus we collect
the coefficients at $r^k$ and they must vanish. We get:
$$\eqalignno{
k=0&\qquad a_0[-l(l+1)]=0\,,\nno{asymk0}\cr
k=1&\qquad a_1[2-l(l+1)]-{2M\over\hbar^2}v_{-1}a_0=0\,,\nno{asymk1}\cr
k\ge2&\qquad a_k[k(k+1)-l(l+1)]-
{2M\over\hbar^2}\sum_{j=0}^{k-1}v_{j-1}a_{k-j-1}+
{2M\over\hbar^2}Ea_{k-2}=0\,.\nno{asymk2}\cr
}$$

This enables us to calculate all the Taylor coefficients of the solution.  To see
how it works, we calculate two examples and compare them to the analytical
solution.
First, let 
$$V=-{Z\over r}\,,$$
so $v_{-1}=-Z$, $v_0=v_1=\dots=0$. For $l=0$, we see from \rno{asymk0} that
$a_0$ can by any number (including zero, but as we will see in a moment, this
would imply the zero solution, which we are obviously not interested in).
From \rno{asymk1} it follows:
$$a_1=-{MZ\over\hbar^2}a_0={-a_0\over a}\,,$$
where $a={\hbar^2\over ZM}$ is the Bohr radius. The first two terms of the
solution are then:
$$R=a_0(1-{r\over a}+O(r^2))\,,$$
which is in agreement with the analytic solution 
\cite{formanek} (eq. 2.524) (every $R_{nl}$ for $l=0$):
$$\eqalign{
R_{10}&=2\sqrt{1\over a^3}\exp\left(-{r\over a}\right)\,,\cr
R_{20}&=\sqrt{1\over2a^3}\left[1-{r\over2a}\right]\exp\left(-{r\over 2a}\right)\,,\cr
R_{30}&={2\over3}\sqrt{1\over3a^3}\left[1-{2r\over3a}+
{2\over27}\left(r\over a\right)^2\right]\exp\left(-{r\over3a}\right)\,,\cr
R_{40}&={1\over4a^{3/2}}\sqrt{1\over a^3}\left[1-{3r\over4a}+
{1\over8}\left(r\over a\right)^2-{1\over192}\left(r\over
a\right)^3\right]\exp\left(-{r\over4a}\right)\,.\cr
}$$

As the second example, we use a linear harmonic oscillator
$$V={M\omega^2r^2\over2}\,,$$
so $v_{-1}=v_0=v_1=0$, $v_2={M\omega^2\over2}$, $v_3=v_4=\dots=0$. 
For $l=0$, we see from \rno{asymk0} that $a_0$ is any number, from 
\rno{asymk1} it follows $a_1=0$ and from \rno{asymk2} we get
$a_2=-{MEa_0\over3\hbar^2}$. But we know the spectrum 
\cite{formanek} (eq. 2.484):
$$E=\hbar\omega(2n+l+{3\over2})\,,$$
so we have $a_2=-{M\omega(2n+{3\over2})a_0\over3\hbar}=-({2\over3}n+{1\over2})
{a_0\over a^2}$, where we used the substitution $a=\sqrt{\hbar\over M\omega}$.
Finally the first two nonzero terms of the solution are:
$$R=a_0\left(1-\left({2\over3}n+{1\over2}\right){r^2\over a^2}+
O(r^3)\right)\,,$$
which agrees with the analytic solution \cite{formanek} (eq. 2.488)
(again every $R_{nl}$ for $l=0$):
$$\eqalign{
R_{00}&={2\over\pi^{1/4}}\sqrt{1\over3a^3}\exp\left(-{r^2\over2a^2}\right)\,,\cr
R_{10}&={1\over\pi^{1/4}}\sqrt{6\over a^3}\left[1-
{2\over3}\left(r\over a\right)^2\right]\exp\left(-{r^2\over2a^2}\right)\,.\cr
}$$

These examples show, that for $l=0$ the derivative $R'$ (the second term in
the $R$ expansion) nontrivially depends on $V$ in the first example, and on $E$ in the second
example. Which is inconvenient for a numerical computation.

For $l>0$, the Taylor coefficients can be calculated in the same way as 
for $l=0$.
From \rno{asymk0}, \rno{asymk1} and \rno{asymk2} we see that 
$a_k=0$ for all $k<l$. So indeed the first nonzero term is $a_lr^l$ as
expected.

%% file: chapter2.tex
\chapter{Dirac Equation}

\section{Introduction}

The Dirac equation for one particle is \cite{strange,zabloudil}:
$$H\psi=W\psi\,,\no{diraceq}$$
$$H=c\balpha\cdot{\bf p}+\beta mc^2+V(r)\hbox{\dsrom1}\,,$$
where $\psi$ is a four component vector:
$$\psi=\left(\matrix{\psi_1\cr\psi_2\cr\psi_3\cr\psi_4\cr}\right)
=\col{\psi_A}{\psi_B}\,,\qquad
\psi_A=\col{\psi_1}{\psi_2}\!,\,\psi_B=\col{\psi_3}{\psi_4}$$
and $\balpha$, $\beta$ are $4\times4$ matrices:
$$\balpha=\mat{0}{\bsigma}{\bsigma}{0}\,,$$
$$\beta=\mat{\hbox{\dsrom1}}{0}{0}{-\hbox{\dsrom1}}\,,$$
where the Pauli matrices $\bsigma=(\sigma_x,\sigma_y,\sigma_z)$ and
$\hbox{\dsrom1}$ form a basis of all $2\times2$ Hermitian matrices.
Substituting all of this into \rno{diraceq} yields: 
$$c\,\bsigma\cdot{\bf p}\col{\psi_B}{\psi_A}=
\matd{W-V-mc^2}{W-V+mc^2}\col{\psi_A}{\psi_B}\,.\no{dirac2}$$
To derive a continuity equation, we multiply \rno{diraceq} by $\psi^*$
and subtract the conjugate transpose of \rno{diraceq} multiplied by $\psi$:
$$\p{}{t}(\psi^*\psi)=-\nabla\cdot(c\psi^*\balpha\psi)\,,$$
so we identify the probability and current densities as
$$\rho=\psi^*\psi=\psi_1^*\psi_1+\psi_2^*\psi_2+\psi_3^*\psi_3+\psi_4^*\psi_4\,,
\qquad {\bf j}=c\psi^*\balpha\psi\,.$$
The normalization of a four-component wave function is then
$$
\int \rho \,\d^3x=
\int \psi^*\psi \,\d^3x=
\int \psi_1^*\psi_1+\psi_2^*\psi_2+\psi_3^*\psi_3+\psi_4^*\psi_4 \,\d^3x=
1\,.\no{norm}$$
The probability density $\rho(x,y,z)$ is the physical quantity we are
interested in, and all the four-component wavefunctions and other formalism is
just a way of calculating it. This $\rho$ is also the thing we should compare
with the Schr\"odinger equation. 

\section{Derivation of the radial equation}

We rewrite $\bsigma\cdot{\bf p}$ \cite{branson}:
$$\bsigma\cdot{\bf p}={1\over r}{\bsigma\cdot{\bf x}\over r}
\left(-i\hbar r \p{}{r}+i\bsigma\cdot{\bf L}\right),\no{sigmap}$$
and search for a basis in the form
$$\psi_A=g\chi^{j_3}_{\kappa}\,,\no{psia}$$
$$\psi_B=if\chi^{j_3}_{-\kappa}\,.\no{psib}$$
We chose this form, because we want $\psi_A$ and $\psi_B$ to have the same $j$
and $j_3$, so the only thing they can differ in is $l$ (the two allowed
possibilities are $l=j\pm\half$). According to \rno{chi1}, \rno{chi2},
\rno{chi3} and \rno{chi2}, this corresponds to $\pm\kappa$ (see also
\cite{kellog,strange,rose,branson} and the appendix for more information about
spin-angular functions).

Substituting \rno{psia}, \rno{psib} and \rno{sigmap} into \rno{dirac2} we get:
$$c{1\over r}{\bsigma\cdot{\bf x}\over r} \left(-i\hbar r
\p{}{r}+i\bsigma\cdot{\bf
L}\right)\col{if\chi^{j_3}_{-\kappa}}{g\chi^{j_3}_{\kappa}}=
\hskip3cm
$$ 
$$
\hskip3cm
=
\matd{W-V-mc^2}{W-V+mc^2}\col{g\chi^{j_3}_{\kappa}}{if\chi^{j_3}_{-\kappa}}\,,$$
Both $\psi_A$ and $\psi_B$ are eigenvectors of the operator 
$K=\beta(\bsigma\cdot{\bf L}+\hbar)$:
$$K\psi_{A,B}=-\hbar\kappa\psi_{A,B}\,,$$
so the action of the $\bsigma\cdot{\bf L}$ operator on the
$\psi_A$ and $\psi_B$ can easily be determined:
$$\bsigma\cdot{\bf L}\col{\psi_A}{\psi_B}=
(\beta K-\hbar)\col{\psi_A}{\psi_B}=
\hskip3cm
$$ 
$$
\hskip3cm
=
\matd{K-\hbar}{-K-\hbar}\col{\psi_A}{\psi_B}=
\col{\hbar(-\kappa-1)\psi_A}{\hbar(\kappa-1)\psi_B}$$
and we can write
$$c{1\over r}{\bsigma\cdot{\bf x}\over r} 
\col{(-i\hbar r\p{}{r}+i(\kappa-1)\hbar)if\chi^{j_3}_{-\kappa}}
{(-i\hbar r\p{}{r}+i(-\kappa-1)\hbar)g\chi^{j_3}_{\kappa}}=
\hskip3cm
$$ 
$$
\hskip3cm
=
\matd{W-V-mc^2}{W-V+mc^2}\col{g\chi^{j_3}_{\kappa}}{if\chi^{j_3}_{-\kappa}}\,.$$
The operator ${\bsigmasmall\cdot{\bf x}\over r}$ also only acts on the angular
momentum parts of the state \cite{strange} (page 59). 
From \rno{parity:appendix}:
$${\bsigma\cdot{\bf x}\over r}\chi^{j_3}_{\kappa}=-\chi^{j_3}_{-\kappa}\,,$$
$${\bsigma\cdot{\bf x}\over r}\chi^{j_3}_{-\kappa}=-\chi^{j_3}_{\kappa}\,,$$
so
$$c{1\over r} 
\col{(i\hbar r\p{}{r}-i(\kappa-1)\hbar)if\chi^{j_3}_{\kappa}}
{(i\hbar r\p{}{r}+i(\kappa+1)\hbar)g\chi^{j_3}_{-\kappa}}=
\hskip3cm
$$ 
$$
\hskip3cm
=
\matd{W-V-mc^2}{W-V+mc^2}\col{g\chi^{j_3}_{\kappa}}{if\chi^{j_3}_{-\kappa}}\,,$$
rewriting
$$\hbar c{1\over r} 
\matd{(-r\p{}{r}+(\kappa-1))f}
{(r\p{}{r}+(\kappa+1))g}\col{\chi^{j_3}_{\kappa}}{i\chi^{j_3}_{-\kappa}}=
\hskip3cm$$
$$\hskip3cm=\matd{(W-V-mc^2)g}{(W-V+mc^2)f}
\col{\chi^{j_3}_{\kappa}}{i\chi^{j_3}_{-\kappa}}\,,$$
and canceling the same terms on both sides finally yields
$$\hbar c
\col{-\p{f}{r}+{\kappa-1\over r}f}
{\p{g}{r}+{\kappa+1\over r}g}=
\col{(W-V-mc^2)g}{(W-V+mc^2)f}\,.\no{radialdirac}$$
This is the radial Dirac equation. As we shall see in the next section, the
equation for $g$ is (with the exception of a few relativistic corrections)
identical to the radial Schr\"odinger equation. And $f$ vanishes
in the limit $c\to\infty$. For this reason $f$ is called the small 
(fein, minor) component and $g$ the large (gro\ss, major) component. 

The probability density is
$$\rho=\psi^*\psi=\psi^*_A\psi_A+\psi^*_B\psi_B=
f^2\chi^{j_3*}_{-\kappa}\chi^{j_3}_{-\kappa}+
g^2\chi^{j_3*}_{\kappa}\chi^{j_3}_{\kappa}\,,
$$
so from the normalization condition \rno{norm} we get
$$
\int \rho \,\d^3x=
\int f^2\chi^{j_3*}_{-\kappa}\chi^{j_3}_{-\kappa}+
g^2\chi^{j_3*}_{\kappa}\chi^{j_3}_{\kappa} \,\d^3x=
\int (f^2\chi^{j_3*}_{-\kappa}\chi^{j_3}_{-\kappa}+
g^2\chi^{j_3*}_{\kappa}\chi^{j_3}_{\kappa})\,r^2\,\d r\d\Omega=
$$
$$
=\int_0^\infty\!\!\!\!\!f^2r^2\,\d r\int\chi^{j_3*}_{-\kappa}\chi^{j_3}_{-\kappa}
\,\d\Omega+
\int_0^\infty\!\!\!\!\!g^2r^2\,\d r\int\chi^{j_3*}_{\kappa}\chi^{j_3}_{\kappa}
\,\d\Omega=
\int_0^\infty r^2(f^2+g^2)\,\d r=1\,,
$$
where we used the normalization of spin-angular functions \rno{spinnorm}.
Also it can be seen, that the radial probability density
is 
$$\rho(r)=r^2(f^2+g^2)$$
(i.e., the probability to find the electron
between $r_1$ and $r_2$ is $\int_{r_1}^{r_2}r^2(f^2+g^2)\,\d r$). In the 
nonrelativistic case, the density is given by
$$\rho(r)=r^2R^2\,,$$
so the correspondence between the Schr\"odinger and Dirac equation is 
$R^2=f^2+g^2$.

\section{Numerical integration for a given $E$}

We rewrite \rno{radialdirac} to a form found in \cite{strange} (eq. 8.10 and
8.12):
$$\eqalignno{
g_\kappa'&=-{\kappa+1\over r}g_\kappa+{1\over c\hbar}(W-V+mc^2)f_k\,,\nno{strange1}\cr
f_\kappa'&=+{\kappa-1\over r}f_\kappa-{1\over c\hbar}(W-V-mc^2)g_k\,.\nno{strange2}\cr
}$$
Let $\hbar=1$ and define $E=W-mc^2$, so that $E$ doesn't contain the electron
rest mass energy. Next we define a relativistic mass
$$M=m+{1\over2c^2}(E-V)\no{relmass}$$
and we get 
$$\eqalignno{
g_\kappa'&=-{\kappa+1\over r}g_\kappa+2Mcf_k\,,\nno{gkappa}\cr
f_\kappa'&=+{\kappa-1\over r}f_\kappa+{1\over c}(V-E)g_k\,,\nno{fkappa}\cr
}$$
which are the same equations as in \cite{koelling&harmon} (eq. 2a and 2b).

From \rno{gkappa} we express $f_\kappa$ and substitute it into \rno{fkappa}.
At the same time we introduce a new variable
$$\phi_\kappa={1\over2M}g_\kappa'\,,\no{phikappa}$$
(beware, \cite{koelling&harmon} introduce $\phi_\kappa={1\over2Mc}g_\kappa'$)
to simplify our equations:
$$f_\kappa={g_\kappa'\over2Mc}+{\kappa+1\over r}{g_\kappa\over2Mc}
={\phi_\kappa\over c}+{\kappa+1\over c}{g_\kappa\over2Mr}\,,\no{fk}$$
differentiate
$$f_\kappa'=
{\phi_\kappa'\over c}+{\kappa+1\over c}\left(g_\kappa\over2Mr\right)'
={\phi_\kappa'\over c}+{\kappa+1\over cr}\phi_\kappa+
{\kappa+1\over2M^2cr}g_\kappa{V'\over2c^2}-{\kappa+1\over2Mcr^2}g_\kappa$$
and substitute $f_\kappa$ and $f_\kappa'$ back to \rno{fkappa}:
$$
{\phi_\kappa'\over c}+{\kappa+1\over cr}\phi_\kappa+
{\kappa+1\over2M^2cr}g_\kappa{V'\over2c^2}-{\kappa+1\over2Mcr^2}g_\kappa=$$
$$={\kappa-1\over r}
\left(
{\phi_\kappa\over c}+{\kappa+1\over c}{g_\kappa\over2Mr}
\right)
+{1\over c}(V-E)g_k\,,$$
by a simplification we finally get
$$
\phi_\kappa'(r)=-{2\over r}\phi_\kappa(r)+\left[\big(V(r)-E\big)+
{\kappa(\kappa+1)\over 2M(r)r^2}-
{\kappa+1\over4M^2(r)c^2r}V'(r)\right]g_\kappa(r)\,.\no{phideriv}
$$ 
Equations \rno{phideriv}, \rno{phikappa} and \rno{relmass} are the equations
we are solving. It is instructive to write equation for $g_\kappa$
directly. For this, we need to calculate 
$$\phi_\kappa'={g''\over2M}-{g'M'\over2M^2}$$
and substituting for $\phi_\kappa$ and $\phi_\kappa'$ into \rno{phideriv} gives
$${g_\kappa''\over2M}-{g_\kappa'M'\over2M^2}=-{2\over
r}{g_\kappa'\over2M}+\left[\big(V-E\big)+ {\kappa(\kappa+1)\over 2Mr^2}-
{\kappa+1\over4M^2c^2r}V'\right]g_\kappa\,,$$
multiplying by $2M$ and rewriting
$${g_\kappa''}=-\left({2\over r}-{M'\over M}\right)
g_\kappa'+\left[\big(V-E\big)+ {\kappa(\kappa+1)\over 2Mr^2}-
{\kappa+1\over4M^2c^2r}V'\right]2Mg_\kappa\,.$$
From \rno{relmass} it follows
$${M'\over M}=-{V'\over2Mc^2}\,,$$
so we finally get
$${g_\kappa''}=-\left({2\over r}+{V'\over2Mc^2}\right)
g_\kappa'+\left[\big(V-E\big)+ {\kappa(\kappa+1)\over 2Mr^2}-
{\kappa+1\over4M^2c^2r}V'\right]2Mg_\kappa\,.\no{eqgk}$$
Comparing \rno{eqgk} with \rno{radial} we clearly see the two relativistic
corrections, both depending on $V'$ and both vanishing for $c\to\infty$ as
expected. Also it should be noted that there is another difference, that
\rno{eqgk} contains the relativistic mass \rno{relmass}, but \rno{radial}
contains the rest mass $m$.

All the terms in the \rno{eqgk} are the same for both possibilities
$j=l\pm\half$ (i.e. $\kappa(\kappa+1)=l(l+1)$ for
both $\kappa=-l-1$ and $\kappa=l$), 
except for the spin-orbit coupling
term ${\kappa+1\over4M^2c^2r}V'(r)$, see for example \cite{strange}
(eq. 2.64).
Sometimes it makes sense to
consider a semirelativistic case, where we neglect the
spin-orbit coupling term, in which case we are left with the
Schr\"odinger equation with the relativistic mass $M$ and only one
correction ${V'\over2Mc^2}$. 

In practice, the potential $V$ is given on a discrete grid, so we need to
compute it's derivative $V'$ numerically. This is the reason we solve 
\rno{phideriv} instead of \rno{eqgk}, because we need to evaluate $V'$ only
once for the spin-orbit term (for the semirelativistic case we don't even need
the $V'$ at all). But besides this minor technical thing, there is no other
reason we chose \rno{phideriv} and not \rno{eqgk}, which looks more familiar.

Once we have calculated $g_\kappa$ and $\phi_\kappa$, we calculate $f_\kappa$ 
from \rno{fk}. So the result of the radial
Dirac equation are two functions $f_\kappa$ and $g_\kappa$. The
physically relevant quantity is the radial probability density
$$\rho(r)=r^2(f^2_\kappa+g^2_\kappa)\,.$$

We calculate the functions $f_\kappa$ and $g_\kappa$ in a similar
way as we calculated $R$ for the Schr\"odinger equation, thus we 
need the asymptotic behavior at the origin. The potential
can always be treated as $V=1/r+\cdots$ and in this case 
it can be shown \cite{zabloudil}, that the asymptotic is 
$$g_\kappa=r^{\beta-1}\,,$$
$$\phi_\kappa={(\beta-1)r^{\beta-2}\over2M}\,,$$
where 
$$\beta=\sqrt{\kappa^2-\left(Z\over c\right)^2}\,,\no{diracasymptotic}$$
or, if we write it explicitly, for $j=l+\half$
$$\beta^+=\sqrt{(-l-1)^2-\left(Z\over c\right)^2}$$
and $j=l-\half$
$$\beta^-=\sqrt{l^2-\left(Z\over c\right)^2}\,.$$
In the semirelativistic case (which is an approximation --- we neglect
the spin-orbit coupling term) we choose
$$\beta=\sqrt{\half(|\beta^+|^2+|\beta^-|^2)}=
\sqrt{l^2+l+\half-\left(Z\over c\right)^2}\,.$$
It should be noted that in the literature we can find other types of 
aymptotic behaviour for the semirelativistic case, its just a question of the
used approximation. One can hardly say that some of them are correct and
another is not since the semirelativistic (sometimes denoted as
scalar-relativstic) approximation itself is not correct, it's just an
approximation.

It follows from \rno{diracasymptotic} that for $j=l+\half$ the radial Dirac
equation completely becomes the radial Schr\"odinger equation in the limit
$c\to\infty$ (and gives exactly the same solutions). For $j=l-\half$ however, 
we get a wrong asymptotic: we get a radial Schr\"odinger equation for $l$, but
the asymptotic for $l-1$.

\section{Other forms of the equations}

Unfortunately, there are a lot of different forms the radial Dirac equation can
be found in the literature. Many authors use different symbols, different
units, I even found a mistake in \cite{donald:apw}. For the reader's comfort,
this section is devoted to deriving and presenting the most common forms of the
equations. 

We realize the fact
$$\eqalign{
f_\kappa'-{\kappa-1\over r}f_\kappa&=
{1\over r}\left({\d\over\d r}-{\kappa\over r}\right)(rf_\kappa)\,,\cr
g_\kappa'+{\kappa+1\over r}g_\kappa&=
{1\over r}\left({\d\over\d r}+{\kappa\over r}\right)(rg_\kappa)\,,\cr
}$$
and rewrite equations \rno{strange1} and \rno{strange2}:
$$\eqalign{
{1\over r}\left({\d\over\d r}+{\kappa\over r}\right)(rg_\kappa)
&-{1\over c}(W-V+mc^2)f_k=0\,,\cr
{1\over r}\left({\d\over\d r}-{\kappa\over r}\right)(rf_\kappa)
&+{1\over c}(W-V-mc^2)g_k=0\,.\cr
}$$
These equations could be found in \cite{zabloudil} (eq. 8.10 and 8.9).
Let's make the substitution \cite{donald:apw}
$$\eqalign{
P_\kappa&=rg_\kappa\,,\cr
Q_\kappa&=rf_\kappa\cr
}$$
and write
$$\eqalign{
\left({\d\over\d r}+{\kappa\over r}\right)P_\kappa
&-{1\over c}(W-V+mc^2)Q_k=0\,,\cr
\left({\d\over\d r}-{\kappa\over r}\right)Q_\kappa
&+{1\over c}(W-V-mc^2)P_k=0\,,\cr
}$$
which can be found in \cite{engel} (eq. 3 and 4: they write 
$a_{nlj}(r)\equiv P_\kappa$ and $b_{nlj}(r)\equiv Q_\kappa$) and also
in \cite{strange} (eq. 8.13: he uses
$u_\kappa\equiv P_\kappa$ and $v_\kappa\equiv Q_\kappa$).

Now we use \rno{relmass} and these relations become
$$\eqalign{
W-V+mc^2&=E-V+2mc^2=2Mc^2\,,\cr
W-V-mc^2&=E-V\,,\cr
}$$
to write
$$\eqalign{
{\d P_\kappa\over\d r}&=-{\kappa\over r}P_\kappa+2McQ_k\,,\cr
{\d Q_\kappa\over\d r}&={\kappa\over r}Q_\kappa-{1\over c}(E-V)P_k\,,\cr
}$$
which can be found in \cite{donald:apw} (eq. 2a and 2b) (there is a mistake
there, they forgot to divide by $r$). $2M$ can be written explicitly as
$$2M=\left[{E-V\over c^2}+2m\right]=\left[{E-V\over c^2}+2\right]\rm\,a.u.\,,$$
so we get
$$\eqalign{
{\d P_\kappa\over\d r}&=-{\kappa\over r}P_\kappa+\left[{E-V\over c^2}+2\right]cQ_k\,,\cr
{\d Q_\kappa\over\d r}&={\kappa\over r}Q_\kappa-{1\over c}(E-V)P_k\,,\cr
}$$
which can be found in \cite{zabloudil} (eq. 8.12 and 8.13), where they have one
$c$
hidden in $Q_\kappa=crf_\kappa$ and use Rydberg atomic units, so they have $1$ instead of $2$ in the square bracket.
It can be found in \cite{bachelet} as well, they use Hartree atomic units, but
have a different notation $G_\kappa\equiv P_\kappa$ and $F_\kappa\equiv
Q_\kappa$, also they made a substitution $c={1\over\alpha}$.

Some authors also use $\epsilon\equiv E$.

%% file: chapter3.tex
\chapter{Eigenproblem}

\section{Introduction}

In the previous two chapters, we learned how to calculate the solution of both
radial Schr\"odinger and Dirac equations for a given $E$.  For most of the
energies, however, the solution for $r\to\infty$ exponentially diverges to
$\pm\infty$. Only for the energies equal to eigenvalues, the solution tends
exponentially to zero for $r\to\infty$. The spectrum for bounded states is
discrete, so we label the energies by $n$, starting from $1$. 

We want to find the eigenvalue and eigenfunction for a given $n$ and $l$
(and a spin in the relativistic case).
The algorithm is the same for both nonrelativistic and relativistic case and
is based on two facts, first that the number of nodes (ie. the number of
intersections with the $x$ axis, not counting the one at the origin and in the
infinity) of $R_{nl}$ and $g_\kappa$ is $n-l-1$ and second that the solution
must tend to zero at infinity.

\section{Algorithm for solving the eigenproblem}

We calculate the solution for some (random) energy $E_0$, using the
procedure described above. Then we count the number of nodes 
(for diverging solutions, we don't count the last one) and check, if the
solution is approaching the zero from top or bottom in the infinity. From the
number of nodes and the direction it is approaching the zero it can be
determined
whether the energy $E_0$ is below or above the eigenvalue $E$ belonging to a given
$n$ and $l$. The rest is simple, we find two energies, one below $E$, one above
$E$ and by halving the interval we calculate $E$ with any precision we want.

There are a few technical numerical problems that are unimportant from the
theoretical point of view, but that need to be solved if one attempts to
actually implement this algorithm. One of them is that when we end the
algorithm, because the energy interval is sufficiently small, it doesn't mean
the solution is near zero for the biggest $r$ we are integrating the equation.
Remember, the solution goes exponentially to $\pm\infty$ for every $E$ except
the eigenvalues and because we never find the exact eigenvalue, the solution
will (at some point) diverge from zero. 

Possible solution that we have employed is as follows: 
when the algorithm ends we find the last minimum (which is always near
zero) and trim the solution behind it (set it to zero). 

Another solution of this problem is to end the algorithm not only when the
energy interval is small enough, but also when the solution is sufficiently
near zero for the biggest $r$ we integrate. One would thought (including me at
first :) that in this case we don't need to trim the solution. Mistake.
If the biggest $r$ is big enough, then even the variation of the initial
$E$ by $10^{-15}$ is not enough to push the tail of the solution to zero. It
can actually be above zero by even $50\%$ of the maximum of the
wave function or more, so we would have to trim the solution anyway (not
mentioning that the algorithm never ends, because the solution will never be
close enough to zero to pass the ending condition --- that actually cannot even
be formulated\dots).

The second rather technical problem is how to choose the initial interval of
energies so that the eigenvalue lies inside the interval.

%% file: chapter4.tex
\chapter{Implementation}


\section{Fortran routine}

The original Fortran 77 routine solves the equation \rno{radial}
on a hyperbolic grid \rno{hyp} using a simple polynomial approximation at each
step. The
input is the energy $E$, grid parameters $a_P$, $k_0$ and $j_M$, plus an array
with values of the potential $V(r)$ at the grid points. The output is an array
with values of $R_{El}(r)$ at the grid points. Our task was to replace this
routine (with minimum changes in other parts of the code) with 
the new routine that solves the equations \rno{phideriv}, \rno{phikappa} and
\rno{relmass} on the hyperbolic grid using
the fourth-order Runge-Kutta method.  The input is the same as for the original
routine, plus $Z$. The output are the two functions $f$ and $g$.

This means that in the limit $c\to\infty$ the $g$ component 
of the new routine gives the same results as the original one.

This new routine can switch between the fully relativistic (Dirac) case,
a semirelativistic case (we neglect the spin-orbit coupling term) and a
nonrelativistic case (we neglect the spin-orbit coupling term and use
a nonrelativistic mass, which effectively restores the Schr\"odinger
equation).

\section{Eigenvalue problem}

This routine uses the algorithm described in the preceding chapter. It can use
either the new or the old routine for the integration.

\section{Hyperbolic grid}

The routines solve the equations on a hyperbolic grid, which is defined as
$$r=a_P{\xi\over1-\xi}\,,\no{hyp}$$
where $0\le\xi<1$ is a dimensionless parameter and $a_P$ 
is a scaling (in the same units as $r$), it is the
value of $r$ for $\xi=0.5$.
We take equidistant values for $\xi$
$$\xi={1\over j_M}, {2\over j_M}, \dots, {k_0\over j_M},\quad\quad k_0<j_M\,,
\no{ksi}$$
by substituting \rno{ksi} into \rno{hyp} we get
$$r(j)=a_P{j\over j_M-j},\quad j=1,2,\dots,k_0\,.\no{rj}$$
The following values of the grid parameters has proven successful: $a_P=1$,
$k_0=500$ and $j_M=525$.

Routines accepts the parameters $a_P$, $k_0$ and $j_M$, 
then the array $r(j)$ of a dimension $k_0$, which 
gives the values of $r$ in the points $j=1,2,\dots,k_0$ calculated
using the relation \rno{rj}. The routine can easily calculate all the values
of $r(j)$ from \rno{rj}, or use the values directly from the array
$r(j)$. We use the way that is more readable.


A formula for the inverse transformation can be derived from \rno{rj}:
$$j=j_M{r(j)\over a_P+r(j)}\,.$$




%% file: chapter5.tex
\chapter{Results}

\section{One electron atom calculation}

To check that our program works, we have computed a spectrum
of a one electron atom with these parameters:
$Z=92$, $k_0=2920$, $j_M=3000$, $a_P=1.0$. 
The result can be compared to fig. 8.3 in \cite{strange}.

In the following table, $E_1$ represents eigenvalues calculated from the
nonrelativistic formula
$$E_n=-{Z^2\over2n^2}\,,$$
$E_2$ represents eigenvalues in the nonrelativistic case
(so it should not depend on $l$) and $E_3$ eigenvalues computed
from the radial Dirac equation, so it depends on $l$.
The energies are given in eV.

\noindent
Case $j=l+\half$:

\verbatim
n=1, l=0, E1=-115156.9520 E2=-114712.4604 E3=-131994.5289
n=2, l=0, E1= -28789.2380 E2= -28733.2317 E3= -34151.9041
n=2, l=1, E1= -28789.2380 E2= -28789.2246 E3= -29649.2912
n=3, l=0, E1= -12795.2168 E2= -12778.5942 E3= -14649.7703
n=3, l=1, E1= -12795.2168 E2= -12795.2121 E3= -13307.1396
n=3, l=2, E1= -12795.2168 E2= -12795.2167 E3= -12959.5541
n=4, l=0, E1=  -7197.3095 E2=  -7190.2921 E3=  -8026.0999
n=4, l=1, E1=  -7197.3095 E2=  -7197.3073 E3=  -7466.8873
n=4, l=2, E1=  -7197.3095 E2=  -7197.3094 E3=  -7318.8303
n=4, l=3, E1=  -7197.3095 E2=  -7197.3094 E3=  -7248.7232
|endverbatim
\bigskip
\noindent
Case $j=l-\half$ (for $l=0$ we take $j=\half$):

\verbatim
n=1, l=0, E1=-115156.9520 E2=-114712.4604 E3=-131994.5289
n=2, l=0, E1= -28789.2380 E2= -28733.2317 E3= -34151.9041
n=2, l=1, E1= -28789.2380 E2= -28789.2246 E3= -34226.2202
n=3, l=0, E1= -12795.2168 E2= -12778.5942 E3= -14649.7703
n=3, l=1, E1= -12795.2168 E2= -12795.2121 E3= -14673.2066
n=3, l=2, E1= -12795.2168 E2= -12795.2167 E3= -13307.1884
n=4, l=0, E1=  -7197.3095 E2=  -7190.2921 E3=  -8026.0999
n=4, l=1, E1=  -7197.3095 E2=  -7197.3073 E3=  -8035.9706
n=4, l=2, E1=  -7197.3095 E2=  -7197.3094 E3=  -7466.9097
n=4, l=3, E1=  -7197.3095 E2=  -7197.3094 E3=  -7318.8305
|endverbatim
\bigskip
\noindent
It should be noted that the eigenvalues don't depend on $l$ in the
nonrelativistic case only for exactly the Coulomb potential. In practice, for
potentials screened by other electrons in real atoms, this is not the case.

%% file: conclusion.tex
\chapter{Conclusion}

Radial numeric solution of the Dirac equation has been implemented under
specific conditions for {\em all-electron pseudopotential} generating process
and the corresponding computer code has been debugged.

%% file: appendix.tex
\chapter{Appendix}

\section{Atomic units (au)}

There are two types of atomic units --- Hartree units and Rydberg units.
The term "atomic units" usually means Hartree atomic units and we use it in
this meaning as well. Some authors however \cite{zabloudil} mean Rydberg
units when they write "atomic units"\dots

All (Hartree) atomic units can be derived from this relation:
$$\hbar=m=e=4\pi\epsilon_0=1\,,$$
where $m$ is the mass of the electron, $e$ a charge of the electron,
$\epsilon_0$ the permittivity of vacuum.

Examples: dimension of length is a Bohr radius
$$a_0={4\pi\epsilon_0\hbar^2\over me^2}={\hbar\over mc\alpha}=
1{\rm\,a.u.}=0.529{\rm\,\AA}=0.529\cdot10^{-10}\rm\,m\,,$$
where $\alpha$ is the dimensionless fine-structure constant
$$\alpha={e^2\over4\pi\epsilon_0\hbar c}={1\over137.036}\,,$$
so the speed of light is 
$$c={1\over\alpha}{\rm\,a.u.}=137.036\rm\,a.u.\,.$$
Energy is measured in Hartrees, one Hartree is
$$E_H={\hbar^2\over ma_0^2}=mc^2\alpha^2=1{\rm\,a.u.}=27.211\rm\,eV\,.$$
Hydrogen atom Hamiltonian in SI units:
$$H=-{\hbar^2\over2m}\nabla^2-{1\over4\pi\epsilon_0}{e^2\over r}$$
and au units:
$$H=-{1\over2}\nabla^2-{1\over r}\,.$$
The energy spectrum for hydrogen is
$$E_n=-{\hbar^2\over ma_0^2}{1\over2n^2}=-{1\over2n^2}\,.$$

The Rydberg atomic units are defined as
$$\hbar=4\pi\epsilon_0=1\,,\qquad m={1\over2}\,,\qquad e^2=2\,,$$
in these units
$$\eqalign{
a_0&=1\,,\cr
E_H&={\hbar^2\over ma_0^2}=mc^2\alpha^2={1\over2}\,,\cr
E_{Ry}&={\hbar^2\over 2ma_0^2}=1=13.606\rm\,eV\,,\cr
H&=-\nabla^2-{2\over r}\,,\cr
E_n&=-{\hbar^2\over ma_0^2}{1\over2n^2}=-{1\over n^2}\,,\cr
c&={2\over\alpha}=274.072\,.\cr
}$$
This is why the asymptotic term $Z\alpha$ is mostly written as $Z\over c$
(Hartree units), but sometimes as $2Z\over c$ (Rydberg units).  

\section{Spin-angular functions}

The operators $L^2$ (quadrate of angular momentum), $L_3$ ($z$ component of
angular momentum) and $S_3$ ($z$ component of a spin, $\pm\half$) mutually commute, so let's denote the corresponding normalized eigenvectors by
$\ket{lms_3}$. We can however also use commuting operators $J^2$
(quadrate of total momentum), $L^2$ and $J_3$ ($z$ component of total momentum)
and denote their normalized eigenvectors by $\ket{jlj_3}$. 

First we want to find a relation between these two bases. A standard procedure
is to use Clebsch-Gordan coefficients \cite{formanek2}:
$$\ket{jlj_3}=\sum_{m=-l}^l\sum_{s_3=-{1\over2}}^{1\over2}
\cg{l}\half{m}{s_3}{j}{j_3}\,\ket{lms_3}=\no{jlj3}$$
$$=\cg{l}\half{j_3-\half}\half{j}{j_3}\,\ket{l;j_3-\half;\half}+
\cg{l}\half{j_3+\half}{-\half}{j}{j_3}\,\ket{l;j_3+\half;-\half}\,,$$
where $\cg{j_1}{j_2}{m_1}{m_2}{j}{m}$ are Clebsch-Gordan coefficients and we
used the fact that they are nonzero only for $m_1+m_2=m$. Because
we are adding angular momenta $l$ and $\half$, the only possible
values of $j$ are $j=l\pm\half$. $j=l+\half$ means, that
the spin has the same orientation as the angular momentum, $j=l-\half$
corresponds to the opposite orientation. So we will only need these 4
coefficients (see eq. K58 in \cite{formanek} or table 2.2 in \cite{strange}):
$$\eqalign{
\cg{l}\half{j_3-\half}\half{l+\half}{j_3}&=\sqrt{l+j_3+\half\over2l+1}\,,\cr
\cg{l}\half{j_3+\half}{-\half}{l+\half}{j_3}&=\sqrt{l-j_3+\half\over2l+1}\,,\cr
\cg{l}\half{j_3-\half}\half{l-\half}{j_3}&=-\sqrt{l-j_3+\half\over2l+1}\,,\cr
\cg{l}\half{j_3+\half}{-\half}{l-\half}{j_3}&=\sqrt{l+j_3+\half\over2l+1}\,.\cr
}$$
Substituting into \rno{jlj3} yields the desired relation between the two basis:
$$\ket{j=l+\half;l;j_3}=
\sqrt{l+j_3+\half\over2l+1}\,\ket{l;m=j_3-\half;\half}+
\hskip3cm
$$
$$
\hskip3cm
+\sqrt{l-j_3+\half\over2l+1}\,\ket{l;m=j_3+\half;-\half}\,,$$
$$\ket{j=l-\half;l;j_3}=
-\sqrt{l-j_3+\half\over2l+1}\,\ket{l;m=j_3-\half;\half}+
\hskip3cm
$$
$$
\hskip3cm
+\sqrt{l+j_3+\half\over2l+1}\,\ket{l;m=j_3+\half;-\half}\,.$$

Now we want to find representations of these "kets". Our
Hilbert space can be decomposed as 
$$H=\hbox{\dsrom S}^2\otimes\hbox{\dsrom C}^2\,,$$
where $\hbox{\dsrom S}^2$ is a vector space of quadratic integrable functions
on a unit sphere (the angular momentum operator $L$ acts here) and
$\hbox{\dsrom C}^2$ is a two dimensional complex vector space (the spin
operator $S$ acts here). 

The eigenvectors of the $S_3$ operator form a basis of the space $\hbox{\dsrom C}^2$, let's denote them 
$\Phi_s$, where $s=\pm\half$:
$$\Phi_{1\over2}=\col{1}{0}\,,\qquad
\Phi_{-{1\over2}}=\col{0}{1}\,.$$
The eigenvectors of $L^2$ and $L_3$ form a basis of $\hbox{\dsrom S}^2$,
they are called spherical harmonics $Y_{lm}$.

Then
$$\ket{lm;s_3=\half}=Y_{lm}\Phi_{1\over2}=\col{Y_{lm}}0$$
$$\ket{lm;s_3=-\half}=Y_{lm}\Phi_{-{1\over2}}=\col0{Y_{lm}}$$
$$\ket{j=l+\half;lj_3}={1\over\sqrt{2l+1}}
\left(\sqrt{l+j_3+\half}\,Y_{l,m=j_3-{1\over2}}\Phi_{1\over2}+\right.$$
$$\hskip4cm\left.+\sqrt{l-j_3+\half}\,Y_{l,m=j_3+{1\over2}}\Phi_{-{1\over2}}\right)=$$
$$={1\over\sqrt{2l+1}}\col{\sqrt{l+j_3+\half}\,Y_{l,m=j_3-{1\over2}}}
{\sqrt{l-j_3+\half}\,Y_{l,m=j_3+{1\over2}}}=y^{j_3}_{jl}$$
$$\ket{j=l-\half;lj_3}={1\over\sqrt{2l+1}}
\left(-\sqrt{l-j_3+\half}\,Y_{l,m=j_3-{1\over2}}\Phi_{1\over2}+\right.$$
$$\hskip4cm\left.+\sqrt{l+j_3+\half}\,Y_{l,m=j_3+{1\over2}}\Phi_{-{1\over2}}\right)=$$
$$={1\over\sqrt{2l+1}}\col{-\sqrt{l-j_3+\half}\,Y_{l,m=j_3-{1\over2}}}
{\sqrt{l+j_3+\half}\,Y_{l,m=j_3+{1\over2}}}=y^{j_3}_{jl}$$
The $y_{jl}^{j_3}$ are called "spin-angular functions"\cite{sakurai},
or "spin spherical harmonics"\cite{zabloudil}. For the two possibilities
$j=l\pm\half$ we introduce a new quantity $\kappa$:
$$\kappa=\cases{-l-1, &for $j=l+\half$;\cr
    l, &for $j=l-\half$,\cr}$$
so that we can deduce from $\kappa$ both $l$ and the spin orientation and thus
we can label all the spin-angular functions by just two parameters
$\kappa$ and $j_3$ and write:
$$\chi_\kappa^{j_3}\equiv y^{j_3}_{jl}=\col
{\cg{l}\half{j_3-\half}{+\half}{j}{j_3}Y_{l,m=j_3-\half}}
{\cg{l}\half{j_3+\half}{-\half}{j}{j_3}Y_{l,m=j_3+\half}}\,.\no{yjlm}
$$
Any integer value of $\kappa$ is permissible except $\kappa=0$. Some authors
also use a different notation $j_3\equiv m_j\equiv\mu$ for the eigenvalue of
$J_3$.

It can be easily shown, that because of the normalization of spherical
harmonics $Y_{lm}$:
$$\int Y_{lm}^*({\bf n})Y_{l'm'}({\bf n})\d\Omega=\delta_{ll'}\delta_{mm'}\,,$$
the spin-angular functions have the following normalization:
$$\int \chi_\kappa^{j_3*}({\bf n})\chi_{\kappa'}^{j_3'}({\bf
n})\d\Omega=\delta_{\kappa\kappa'}\delta_{j_3j_3'}\,.\no{spinnorm}$$

To be more precise, $-\hbar\kappa$ is actually defined as the eigenvalue of
the operator \cite{strange} (eq. 2.85 and 2.86)
$$K=
\bsigma\cdot{\bf L}+\hbar=
\left({2\over\hbar^2}{\bf S}\cdot{\bf L}+1\right)\hbar=
\left({1\over\hbar^2}({\bf J}^2-{\bf L}^2-{\bf S}^2)+1\right)\hbar\,,
$$
$$\eqalign{K\psi&=\
\left({1\over\hbar^2}({\bf J}^2-{\bf L}^2-{\bf S}^2)+1\right)\hbar\psi=\cr
&=\left(j(j+1)-l(l+1)-s(s+1)+1\right)\hbar\psi=\cr
&=\left(j(j+1)-l(l+1)+{\textstyle{1\over4}}\right)\hbar\psi=\cr
&=-\hbar\kappa\psi\,,\cr
}$$
so we have $\kappa=-j(j+1)+l(l+1)-{\textstyle{1\over4}}$.
For $j=l+\half$: 
$$\kappa=-(l+\half)(l+\half+1)+l(l+1)-{\textstyle{1\over4}}=-l-1$$
and for the other case $j=l-\half$:
$$\kappa=-(l-\half)(l-\half+1)+l(l+1)-{\textstyle{1\over4}}=l\,.$$
Another way of looking at $\kappa$ follows from
$$\kappa=\cases{-l-1=-(j+\half), &for $j=l+\half$;\cr
    l=+(j+\half), &for $j=l-\half$,\cr}$$
which means that $|\kappa|$ is actually the value of the total angular momentum
(plus $\half$ so that we only have integer values) and the sign depends on
whether the spin and the orbital angular momentum have the same or opposite
orientation ($j=l\pm\half$).

In the text, we need to find how the operator
$${\bsigma\cdot{\bf x}\over r}$$
acts on the spin-angular functions. It commutes with ${\bf J}$,
as is easy to prove, which means it also commutes with $J_3$ and ${\bf J}^2$
and so ${\bsigmasmall\cdot{\bf x}\over r}\chi^{j_3}_{\kappa}$ must
have the same values of $j$ and $j_3$ as $\chi^{j_3}_{\kappa}$. Furthermore, it
anticommutes with $K$ \cite{strange}, so
$$K{\bsigma\cdot{\bf x}\over r}\chi^{j_3}_{\kappa}=
-{\bsigma\cdot{\bf x}\over r}K\chi^{j_3}_{\kappa}=
-{\bsigma\cdot{\bf x}\over r}(-\hbar\kappa)\chi^{j_3}_{\kappa}=
\hbar\kappa{\bsigma\cdot{\bf x}\over r}\chi^{j_3}_{\kappa}\,,$$
but also
$$K\chi^{j_3}_{-\kappa}=\hbar\kappa\chi^{j_3}_{-\kappa}\,,$$
which shows that 
$${\bsigma\cdot{\bf x}\over r}\chi^{j_3}_{\kappa}=C\chi^{j_3}_{-\kappa}\,
\no{ceq}$$
for some (complex) constant $C$. From now on, there are mistakes
in the literature. For example \cite{strange} implicitly concludes at this
point that $C=-1$, which is obviously incorrect. It can easily be shown that
$({\bsigmasmall\cdot{\bf x}\over r})^2=\hbox{\dsrom 1}$, so $C^2=1$,
or $C=\pm1$. The sign of $C$ can generally depend on 
$\kappa$ and $j_3$, but doesn't depend on ${\bf x}$. To determine it,
one option is to find it explicitly for all values of $\kappa$ and $j_3$.
\cite{rose} chooses ${\bf x}={\bf\hat e}_z$, the unit vector along the
$z$-axis, and writes
$$Y_{lm}({\bf\hat e}_z)=\sqrt{2l+1\over4\pi}\,\delta_{m0}\,,\no{Yz}$$
$${\bsigma\cdot{\bf x}\over r}=\sigma_z\,,$$
so using \rno{ceq}
$$\sigma_z\chi^{j_3}_\kappa=\sigma_zy^{j_3}_{j,j\pm\half}=C\chi^{j_3}_{-\kappa}
=Cy^{j_3}_{j,j\mp\half}\,,$$
where the $+$ sign corresponds to $\kappa>0$ and the $-$ sign to $\kappa<0$.
Using \rno{yjlm} 
$$\col
{\cg{j\pm\half}\half{j_3-\half}{+\half}{j}{j_3}Y_{j\pm\half,j_3-\half}}
{-\cg{j\pm\half}\half{j_3+\half}{-\half}{j}{j_3}Y_{j\pm\half,j_3+\half}}
=
\hskip2cm
$$
$$
\hskip2cm
=C\col
{\cg{j\mp\half}\half{j_3-\half}{+\half}{j}{j_3}Y_{j\mp\half,j_3-\half}}
{\cg{j\mp\half}\half{j_3+\half}{-\half}{j}{j_3}Y_{j\mp\half,j_3+\half}}\,,
$$
evaluating the Clebsch-Gordan coefficients
$${1\over\sqrt{2j+1\pm1}}
\col
{\mp\sqrt{j\mp j_3+\half\pm\half}Y_{j\pm\half,j_3-\half}}
{-\sqrt{j\pm j_3+\half\pm\half}Y_{j\pm\half,j_3+\half}}
=
\hskip2cm
$$
$$
\hskip2cm
=
{1\over\sqrt{2j+1\mp1}}
C
\col
{\pm\sqrt{j\pm j_3+\half\mp\half}Y_{j\mp\half,j_3-\half}}
{\sqrt{j\mp j_3+\half\mp\half}Y_{j\mp\half,j_3+\half}}\,.
$$
Using \rno{Yz} it follows that $C=-1$ for
$j>0$ and $j_3=\pm\half$ (and both $\pm$). For other values
of $j_3$ however, the spherical harmonic $Y_{lm}({\bf\hat e}_z)=0$
(because $m\neq0$), so we can say nothing. Unfortunately, $j_3$ can be any
number from $-j,-j+1,\dots,j$, so proving that $C=-1$ only for
$j_3=\pm\half$ is insufficient and \cite{rose} is wrong as well. We'll give a
correct proof using a Wigner-Eckart theorem that $C=-1$ indeed in 
a moment. So we finally get
$${\bsigma\cdot{\bf x}\over r}\chi^{j_3}_{\kappa}=-\chi^{j_3}_{-\kappa}\,.
\no{parity:appendix}$$
Let's look at the relation between $\chi_\kappa^{j_3}$ and
$\chi_{-\kappa}^{j_3}$ more closely. For $\kappa\ge0$:
$$\eqalignno{
\chi_\kappa^{j_3}&=y_{\kappa-{1\over2},\,\kappa}^{j_3}\,,\nno{chi1}\cr
\chi_{-\kappa}^{j_3}&=y_{\kappa-{1\over2},\,(\kappa-1)}^{j_3}=
y_{(\kappa-1)+{1\over2},\,(\kappa-1)}^{j_3}\,,\nno{chi2}\cr
}$$
for $\kappa<0$:
$$\eqalignno{
\chi_\kappa^{j_3}&=y_{-\kappa-{1\over2},\,(-\kappa-1)}^{j_3}=
y_{(-\kappa-1)+{1\over2},\,(-\kappa-1)}^{j_3}\,,\nno{chi3}\cr
\chi_{-\kappa}^{j_3}&=y_{-\kappa-{1\over2},\,-\kappa}^{j_3}\,.\nno{chi4}\cr
}$$
Basically, $\chi_\kappa^{j_3}$ and $\chi_{-\kappa}^{j_3}$ have
the same $j$ and $j_3$, and differ in $l$. There are only two possibilities
$l=j\pm\half$, so $\chi_\kappa^{j_3}$ picks one and $\chi_{-\kappa}^{j_3}$ has
the other one (which one depends on the sign of $\kappa$). So the operator
${\bsigmasmall\cdot{\bf x}\over r}$ preserves $j$ and $j_3$ and flips 
the $l$ to the other possibility.

Let's get back to determining the constant $C$. We denote the eigenvector
of $J_1^2$, $J_2^2$, $J^2$ and $J_z$ by a ket 
$$\ket{j_1 j_2 j m}\,,$$
where ${\bf J}={\bf J_1}+{\bf J_2}$. In this notation, 
$$\chi^m_\kappa=\braket{{\bf x}|j\pm\half;\half j m}\,,$$
where ${\bf J}_1$ is the orbital angular momentum, ${\bf J}_2$ is the spin,
$j=|\kappa|-\half$ and the plus sign is taken for $\kappa>0$, negative sign is
for $\kappa<0$.
We want to determine the constant $C$ in
$${\bsigma\cdot{\bf x}\over r}\chi^{j_3}_{\kappa}=C\chi^{j_3}_{-\kappa}\,,$$
that is
$${\bsigma\cdot{\bf x}\over r}\ket{j\pm\half;\half j m}=C
\ket{j\mp\half;\half j m}\,.$$
The kets $\ket{j_1 j_2 j m}$ are orthogonal, so 
$$C=\bra{j\mp\half;\half j m}{\bsigma\cdot{\bf x}\over r}\ket{j\pm\half;\half j m}\,.$$
We introduce two vector operators 
$${\bf T}={{\bf x}\over r}\,,$$
$${\bf V}=\bsigma\,,$$
and form tensor operators $T(1,q)$ and $V(1,q)$ ($q=-1,0,1$) in the standard
way:
$$T(1,0)={x_3\over r},\qquad T(1,\pm1)=\mp{x_1\pm i x_2\over r\sqrt{2}}\,,$$
$$V(1,0)=\sigma_3,\qquad V(1,\pm1)=\mp{\sigma_1\pm i\sigma_2\over\sqrt{2}}\,,$$
they obey the following commutation relations
$$[T(1,q),{\bf J}_2]=0\,,$$
$$[V(1,q),{\bf J}_1]=0\,,$$
so we can use the Wigner-Eckart (W-E) theorem for the scalar product of two
tensor operators (note that 
$\hbox{\dsrom T}(1)\cdot\hbox{\dsrom V}(1)\equiv\sum_q(-1)^qT(1,q)V(1,-q)=
T_1V_1+T_2V_2+T_3V_3$): 
$$C=
\braket{j\mp\half;\half j m|{\bsigma\cdot{\bf x}\over r}|j\pm\half;\half j m}=
\braket{j\mp\half;\half j m|
\hbox{\dsrom T}(1)\cdot\hbox{\dsrom V}(1) |j\pm\half;\half j m}=
$$
$$=(-1)^{j+j\pm\half+\half} \jsym{j}{j\mp\half}{\half}{1}{\half}{j\pm\half}
(j\mp\half\,\|\,\hbox{\dsrom T}(1)\,\|\,j\pm\half)
(\half\,\|\,\hbox{\dsrom V}(1)\,\|\half)\,.
$$
Now (remember $j$ can be either integer or half integer)
$$(-1)^{j+j\pm\half+\half}=\mp(-1)^{2j}\,,$$
$$
\jsym{j}{j\mp\half}{\half}{1}{\half}{j\pm\half}=
\jsym{j}{\half}{j+\half}{1}{j-\half}{\half}=
-{(-1)^{2j}\over \sqrt{6(j+\half)}}\,,
$$
$$
(j\mp\half\,\|\,\hbox{\dsrom T}(1)\,\|\,j\pm\half)=\mp\sqrt{j+\half}\,,
$$
$$
(\half\,\|\,\hbox{\dsrom V}(1)\,\|\half)=\sqrt{6}\,,
$$
so
$$C=\mp(-1)^{2j}\left(-{(-1)^{2j}\over \sqrt{6(j+\half)}}\right)
\left(\mp\sqrt{j+\half}\right)\sqrt{6}=-1\,.$$
The $6j$ symbol was determined from the formula \cite{formanek} (K.108)
$$\jsym{a}{b}{c}{1}{c-1}{b}=(-1)^{a+b+c} \sqrt{
(a+b+c+1)(-a+b+c)(a-b+c)
\over
b(2b+1)(2b+2)(2c-1)2c(2c+1)
}\,\cdot$$
$$\cdot\sqrt{a+b-c+1}\,,$$
The reduced matrix element $(\half\,\|\,\hbox{\dsrom V}(1)\,\|\half)$ was
determined from another application of the W-E theorem:
$$ \braket{\half \half | V(1,0) | \half \half}=(-1)^{\half+1-\half}
{\cg{1}{\half}{0}{\half}{\half}{\half}\over\sqrt{2\cdot\half+1}}
(\half\,\|\,\hbox{\dsrom V}(1)\,\|\half)= $$
$$= {1\over\sqrt6}(\half\,\|\,\hbox{\dsrom V}(1)\,\|\half)\,, $$
but also 
$$
\braket{\half \half | V(1,0) | \half \half}=
\braket{\half \half | \sigma_3 | \half \half}=
\row{1}{0}\matd{1}{-1}\col{1}{0}=1\,,
$$
so
$$ (\half\,\|\,\hbox{\dsrom V}(1)\,\|\half)=\sqrt{6}\,. $$
Similarly for the reduced matrix element
$(j\mp\half\,\|\,\hbox{\dsrom T}(1)\,\|\,j\pm\half)$:
$$
\braket{j\mp\half; m | T(1,0) | j\pm\half; m}=(-1)^{j\mp\half+1-j\mp\half}
{\cg{1}{j\pm\half}{0}{m}{j\mp\half}{m}\over\sqrt{2j\mp1+1}}\,\times
$$
$$\times (j\mp\half\,\|\,\hbox{\dsrom T}(1)\,\|\,j\pm\half)= 
\mp{1\over2}\sqrt{(j+\half)^2-m^2\over j(j+1)(j+\half)}
(j\mp\half\,\|\,\hbox{\dsrom T}(1)\,\|\,j\pm\half)\,,
$$
but also 
$$
\braket{j\mp\half; m | T(1,0) | j\pm\half; m}=
\braket{j\mp\half; m | {x_3\over r} | j\pm\half; m}=
$$
$$=
\int Y^*_{j\mp\half;m}(\vartheta,\varphi)\cos\vartheta\, 
Y_{j\pm\half;m}(\vartheta,\varphi)\,\d\Omega=
{1\over2}\sqrt{(j+\half)^2-m^2\over j(j+1)}\,,
$$
so
$$ (j\mp\half\,\|\,\hbox{\dsrom T}(1)\,\|\,j\pm\half)=\mp\sqrt{j+\half}\,.$$

%% file: references.tex
\references


\bibliography{bachelor}            
\bibliographystyle{plain}